\documentclass[a4paper,11pt]{article}
\usepackage[margin=2.5cm]{geometry}
\usepackage[percent]{overpic}
\usepackage{booktabs}

\usepackage{gensymb}
\usepackage[square,numbers,sort&compress]{natbib}
\usepackage{enumitem}
\setlist{nosep} 

\usepackage{siunitx}

\usepackage{lineno}
\usepackage{hyperref}

\title{AMEGO-X: MeV gamma-ray Astronomy in the Multimessenger Era}

\author{\textbf{Henrike Fleischhack} (for the AMEGO-X team)\\ \href{mailto:fleischhack@cua.edu}{fleischhack@cua.edu}}

\begin{document}
\maketitle
\noindent\textbf{Affiliations:}\\
\noindent Catholic University of America, Department of Physics,
  620 Michigan Ave. N.E., Washington, DC 20064, USA\\NASA Goddard Space Flight Center, 
8800 Greenbelt Rd, Greenbelt, MD 20771, USA\\Center for Research and Exploration in Space Science and Technology, 8800 Greenbelt Rd, Greenbelt, MD 20771, USA \\[0.1cm]

\noindent\textbf{Presented at the 37$^{\rm{th}}$ International Cosmic Ray Conference (ICRC 2021)}, July 12th -- 23rd, 2021; Online -- Berlin, Germany

\section*{Abstract}

Recent detections of gravitational wave signals and neutrinos from gamma-ray sources have ushered in the era of multi-messenger astronomy, while highlighting the importance of gamma-ray observations for this emerging field. AMEGO-X, the All-sky Medium Energy Gamma-Ray Observatory eXplorer, is an MeV gamma-ray instrument that will survey the sky in the energy range from hundreds of keV to one GeV with unprecedented sensitivity. AMEGO-X will detect gamma-ray photons both via Compton interactions and pair production processes, bridging the “sensitivity gap” between hard X-rays and high-energy gamma rays. AMEGO-X will provide important contributions to multi-messenger science and time-domain gamma-ray astronomy, studying e.g. high-redshift blazars, which are probable sources of astrophysical neutrinos, and gamma-ray bursts. I will present an overview of the instrument and science program.

\section{Introduction}

Despite recent improvements in many areas of multi-wavelength astronomy, there is a lack of high-resolution, high-sensitivity observations in the MeV gamma-ray range (between 100 keV and tens of MeV). The most sensitive observations to date have been obtained by COMPTEL\citep{1992NASCP3137...85D} aboard the Compton Gamma Ray Observatory, de-orbited in 2000, and SPI\citep{refId0}, the spectrometer on Integral, which has been taking data since 2002. Neither instrument can provide a sensitivity that is comparable to other detectors in the neighboring hard X-ray or HE gamma-ray wavelengths (see Figure \ref{fig:sensi}).

The lack of instruments imaging MeV gamma-ray sources is not due to a lack of interest by astronomers and astrophysicists, but rather due to the technical difficulty of these observations. MeV gamma-ray photons have a small cross section for interacting with matter. Photons between 100 keV and 10 MeV mainly interact via Compton scattering, which makes it difficult to reconstruct the direction of the incoming photon. 

MeV gamma rays are known or expected to be emitted from a range of sources, including line emission from radioactive elements and electron-positron annihilation and continuum emission from the interactions from relativistic particles with their surroundings. There are also new physics processes that might produce MeV gamma-ray emission. 

In recent years, several mission concepts for an all-sky observatory in the MeV energy range have been proposed, such as AMEGO\citep{mcenery2019allsky} and eASTROGAM\citep{De_Angelis_2018}, none of which have secured funding yet. AMEGO-X, the All-sky Medium Energy Gamma-ray Observatory eXplorer, is a new mission concept that is will be submitted to the next call for proposals for MIDEX (Medium-class explorer) missions by NASA. It builds on the AMEGO concept, with a smaller and lighter detector, reduced sensitivity at high energies ($\sim$GeV), reduced energy threshold, and a more focused science program. AMEGO-X will focus on continuum MeV gamma-ray observations that promote multi-wavelength and multi-messenger science, such as the search for counterparts of sources of gravitational waves, cosmic neutrinos, and cosmic rays in general. Like AMEGO and eASTROGAM, AMEGO-X will be a combined Compton-pair telescope, meaning it is designed to detect gamma rays that interact either via Compton scattering or via pair production.

\section{The AMEGO-X Instrument and Detector}

\begin{figure}[bt]
    \centering
    \includegraphics[height=6.6cm]{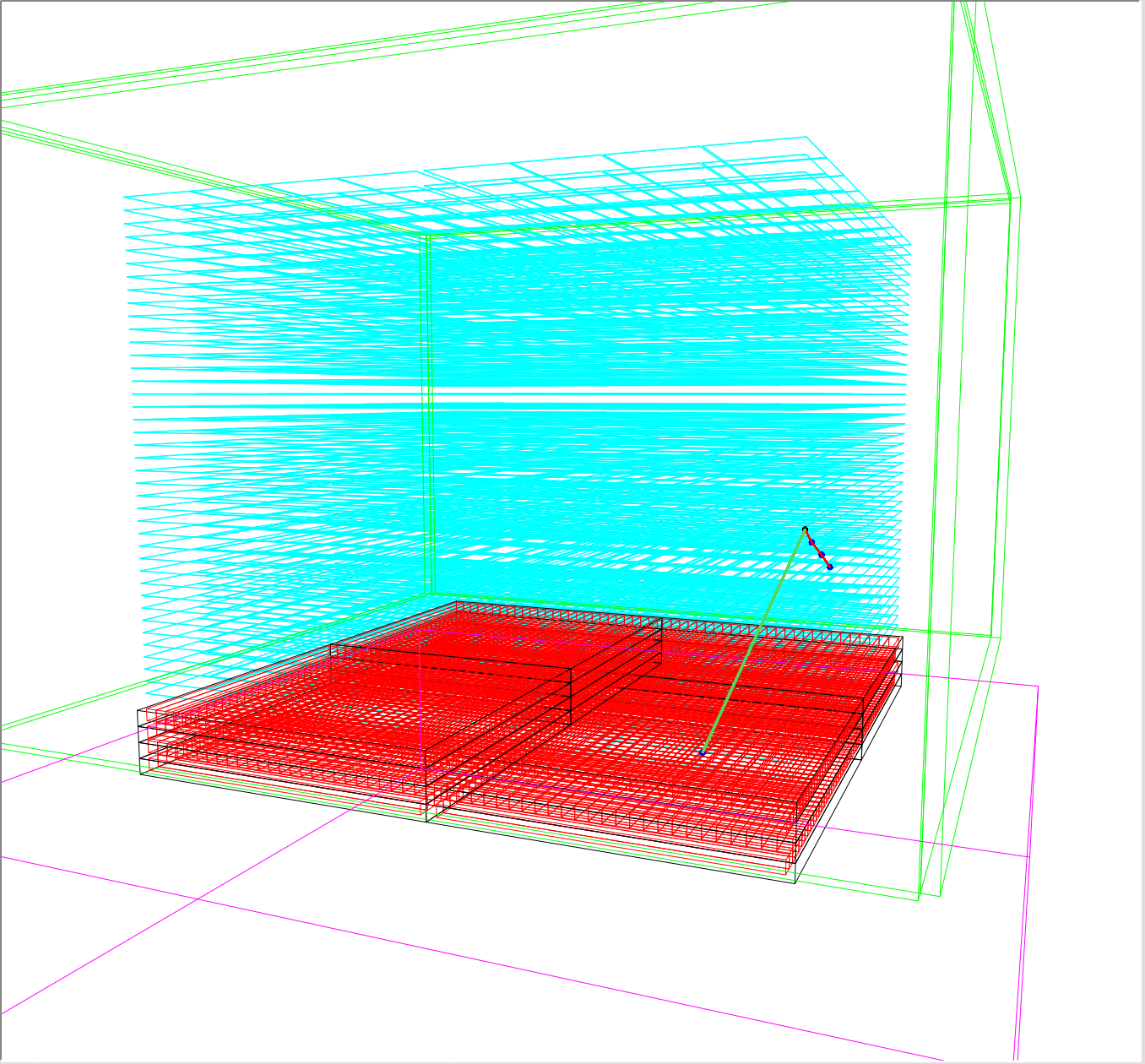} \hspace{0.5cm}
    \includegraphics[height=6.6cm]{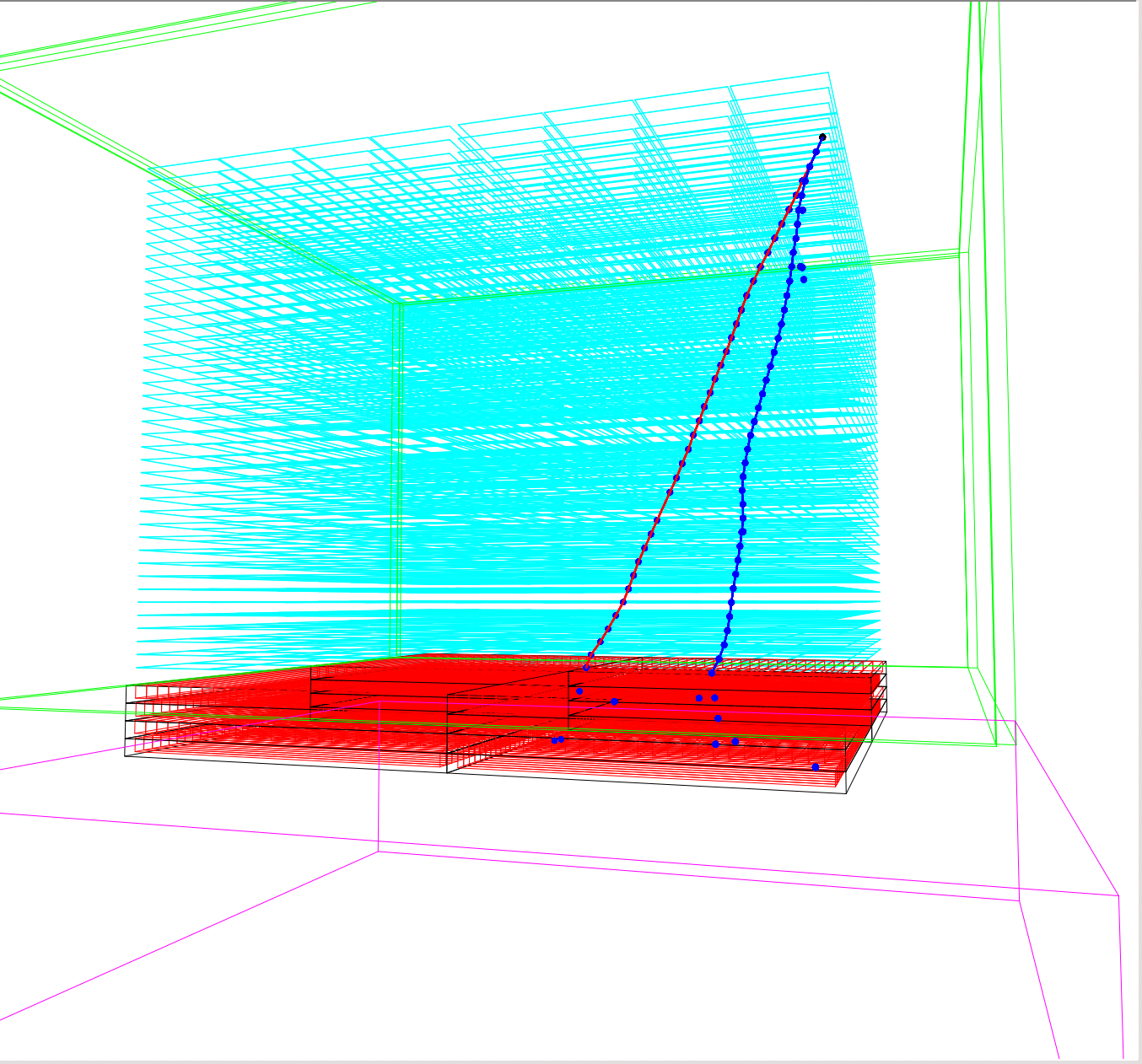}
    \caption{Simulation of a Compton scattering event (left) and pair production event (right) in the AMEGO-X detector. Blue dots represent energy deposits in active detector material. Red line: Reconstructed electron track. Blue line: reconstructed positron track. Pink line: Reconstructed path of scattered photon. The ACD panels are shown in green, the silicon tracker layers in teal, and the CsI calorimeter logs in red. The pink box underneath the detector represents the spacecraft. Passive material (scaffolding, cables etc.) not shown. }
    \label{fig:simevent}
\end{figure}

AMEGO-X will be placed in a low-earth orbit (500 km to 650 km altitude, about 90 minute orbital period) with low inclination ($\leq$5\degree{}). It will use a pointing strategy similar to the original \emph{Fermi}-LAT survey pointing strategy\citep{Atwood_2009}, but with a $\pm$30\degree{} rocking angle. This will provide good coverage of more than 80\% of the sky every day. 

The three main components of the AMEGO-X instrument (see Figure \ref{fig:simevent}) are the AstroPix silicon pixel tracker, the cesium iodide scintillator calorimeter, and the plastic scintillator Anti-Coincidence detector (ACD). To trigger instrument readout, we require at least two hits in the tracker or at least one hit in the tracker and one hit in the calorimeter, with no coincident hit in the ACD. 

\subsection{Tracker}
The tracker (teal in Figure \ref{fig:simevent}) consists of four square towers, each 40 cm wide and 60 cm tall and containing forty layers of the newly developed AstroPix pixel detector chips. These monolithic silicon pixel detectors with integrated readout are based on the ATLASPix pixel design, with a geometry optimized for applications in gamma-ray astronomy\citep{brewer2021astropix, astroPixNim}. The AstroPix pixels are square, with a height of 0.5 mm and a width of 1 mm. Simulations predict comparable performance for pixel width up to 2 mm. The performance results shown here were produced for a slightly older design with a pixel width of 0.5 mm.

The pixelated silicon detectors have a smaller volume and thus lower noise levels compared to the silicon strip detectors commonly used in similar instruments. The low noise level allows us to reduce the trigger threshold in the tracker down to 25 keV. The intrinsic energy resolution of the silicon pixels is about 5\% (Gaussian $\sigma$) at 100 keV.

\subsection{Calorimeter}

The calorimeter (red in Figure \ref{fig:simevent}) is also split into four square towers, each 38 cm wide and 6 cm tall. Each tower is comprised of four layers of 25 cesium-iodide (CsI) scintillator ``logs'', read out on each end by silicon photo-multipliers (SiPMs). The orientation of the logs is offset by 90\degree{} between successive layers to improve the spatial resolution. 

The intrinsic energy resolution of the CsI scintillators is about 7.5\% at 100 keV and 1\% at 5 MeV. The trigger threshold in the calorimeter is 100 keV.

\subsection{ACD}

The ACD consists of five plastic scintillator panels (green in Figure \ref{fig:simevent}) covering the top and sides of the instrument. Each panel is surrounded by wavelength-shifting fibers feeding the collected light into SiPMs. The ACD will be used to veto events with an energy deposit of at least 200 keV in one of the panels, to reduce the background from charged particles (cosmic rays) passing through the instrument. 

\section{Instrument Performance}

\begin{figure}
    \centering
    
    \begin{overpic}[width=\textwidth, trim=0cm 0cm 0cm 1.35cm, clip]{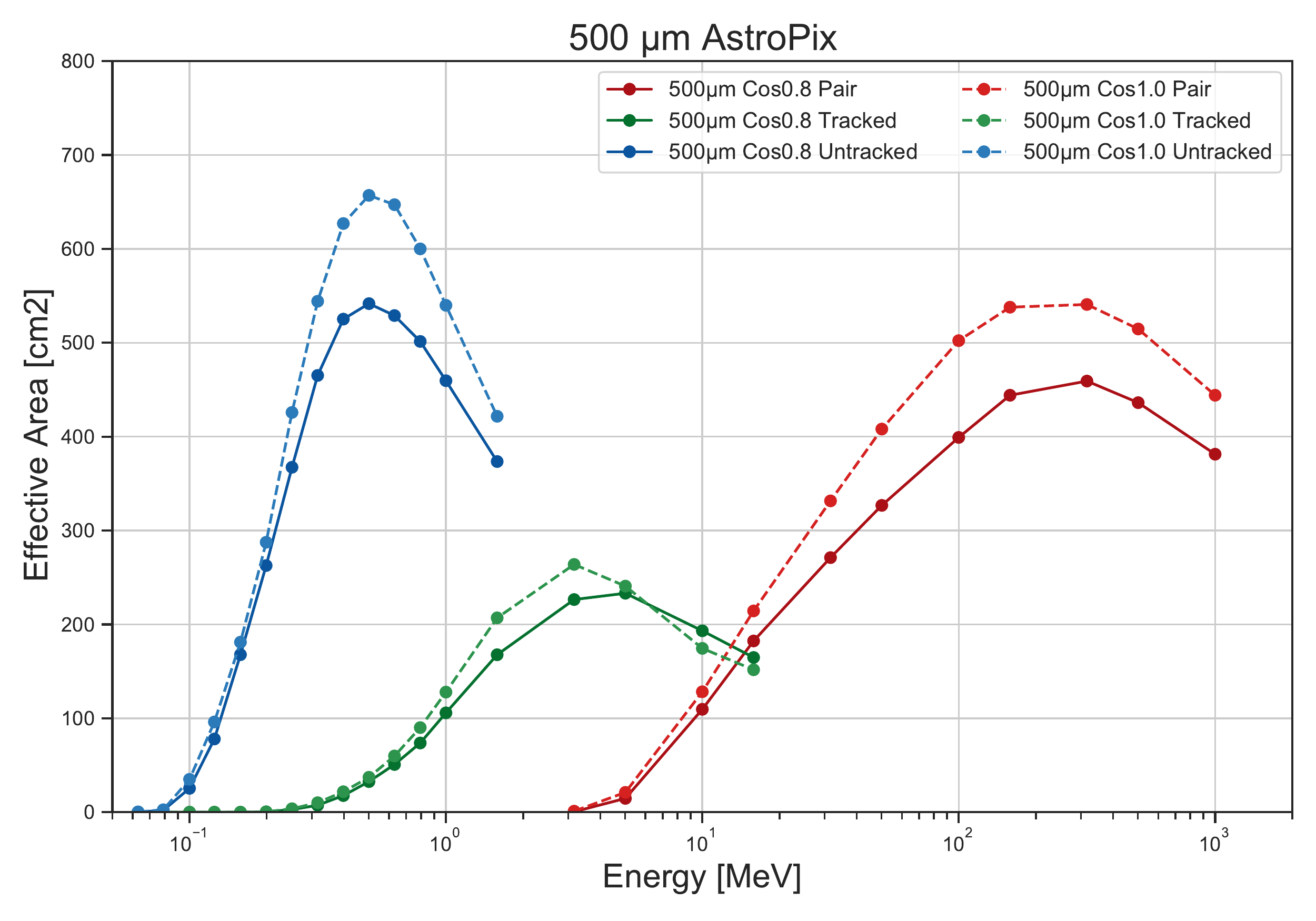}
        \put (10,60) {\textit{Preliminary}}
    \end{overpic}
    ~\\
    \begin{overpic}[width=\textwidth, trim=0cm 0cm 0cm 1.35cm, clip]{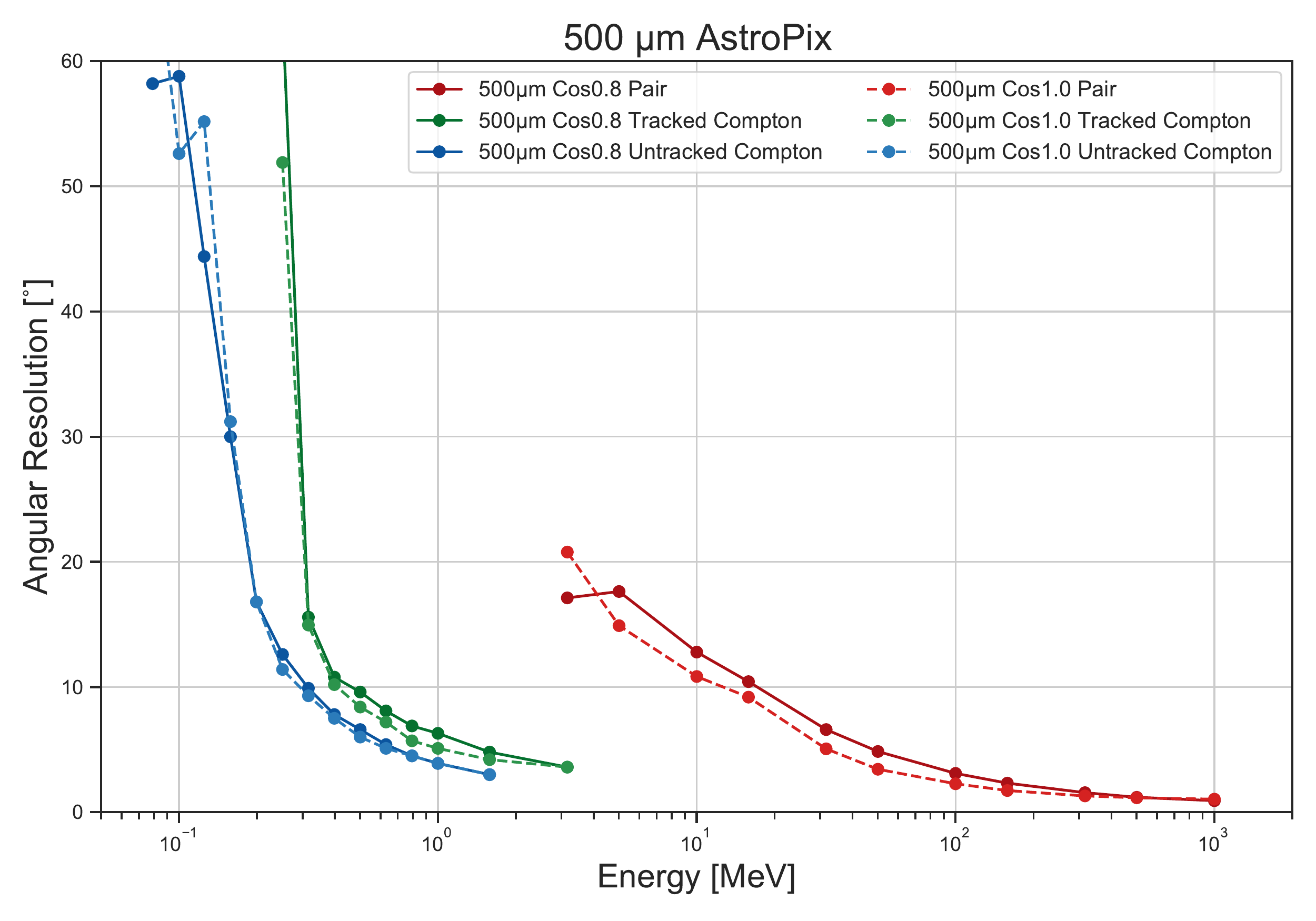}
        \put (10,10) {\textit{Preliminary}}
    \end{overpic}
    \caption{Predicted effective collection area (top) and angular resolution (bottom) for AMEGO-X with the AstroPix tracker, 0.5 mm pixel size. Solid lines are for on-axis gamma-ray photons (``cos1.0'') and dashed lines are for 37\degree{} off-axis photons (``cos0.8'').}
    \label{fig:perfomance}
\end{figure}

The performance of the AMEGO-X instrument was simulated using the GEANT4-based MEGAlib framework \citep{geant,megalib}, which is able to perform all steps of the simulation chain, from generating ensembles of gamma rays and/or cosmic rays with a given beam shape and energy spectrum, simulating passage of these particles through the instrument (including effects like ionization, photo-electric effect, Compton scattering, and the production of secondary particles), simulating energy deposits in active detector material (including electronics noise and energy smearing), triggering, event reconstruction, and the determination of detector performance parameters. 

\subsection{Effective Area}

The effective area is defined as $A_{eff} = \frac{N_{reco}}{N_{sim}}\cdot A_{sim}$, where $N_{reco}$ is the number of reconstructed events of a given event type, $N_{sim}$ is the number of simulated gamma-ray photons at a given (fixed) energy, and $A_{sim}$ is the area of the surface from which the simulated events were thrown. The result is independent of $A_{sim}$ as long as it is big enough to cover the projected geometric area of the instrument. We distinguish three event types: Pair events (with two reconstructed tracks, one from the electron and one from the positron, see Figure \ref{fig:simevent},  right), which dominate above 10 MeV, Compton events with a reconstructed track from the Compton electron (2 MeV to 10 MeV, see Figure \ref{fig:simevent}, left), and Compton events without a track from the Compton electron (below 2 MeV). 

Figure \ref{fig:perfomance} (top) shows the effective area against the true energy for the three different event types. AMEGO-X is sensitive to gamma-ray photons between 100 keV and 1 GeV. Self-vetoes due to energetic electrons and positrons from pair cascades cause a loss of detection efficiency at the highest energies. 

\subsection{Angular Resolution}

For pair events, we are able to reconstruct the directions of the electron and positron from their energy deposits in the tracker. The direction of the initial gamma ray is then estimated as the average of the two, weighted by their energies. The reconstructed direction is generally distributed according to an approximately symmetric distribution around the true direction. We quote the angular resolution for pair events as the 68\% containment radius for $\Delta\Theta$, the angle between the true and reconstructed direction of the initial gamma-ray photon. Figure \ref{fig:perfomance} (bottom) shows the angular resolution versus the simulated energy. For pair events below 10 MeV, the angular resolution tends to be quite poor, 10\degree{} or worse. The angular resolution improves with energy and is generally better than 5\degree{} for energies of 50 MeV or above.

For Compton events, we are typically able to reconstruct the direction of the scattered photon (from the first and second interaction in the tracker and/or calorimeter) and its scattering angle $\theta_C$, given by $\cos\left(\theta_C\right) = 1 - \frac{m_e c^2}{E'_\gamma} - \frac{m_e c^2}{E'_\gamma + E_e}$, where $E'_\gamma$ is the energy of the scattered photon and $E_e$ is the energy of the Compton electron. Thus, the direction of the initial gamma-ray photon is constrained to lay on a circle on the sky with radius $\theta_C$. For events where an electron track can be identified, there is an additional constraint: The initial gamma-ray direction has to lie on the plane defined by the direction of the scattered photon and the Compton electron. 

Due to measurement uncertainties, the reconstructed directions are (in practice) constrained to lie on a \emph{ring} (if no track is present) or a \emph{ring segment} (banana shape) for events with tracks. Since we typically detect many photons from each source, we can use the intersections of these rings and ring segments to fit the source position. Therefore, the ``width'' of the back-projected Cherenkov cone, given by the uncertainty in $\theta_C$, is a good measure for the angular resolution of a Compton telescope. 

Here, we define the ARM (angular resolution measure) for Compton events as the full width half maximum of the distribution of $\Delta \theta_C$, defined as the difference between the Compton scattering angle as derived from the energy measurements as shown above, and the Compton scattering angle as derived from the simulated source position and the positions of the first and second interactions in the tracker. This can also be interpreted as the closest distance of the back-projected Compton ring to the source.  

Figure \ref{fig:perfomance} (bottom) shows the ARM for Compton events versus the simulated energy. For Compton events without track below about 200 keV, and for Compton events with an electron track below 300 keV, the uncertainties are large enough to make imaging analyses challenging. However, for larger energies, the resolution improves and hits about 5\degree{} at the 1 MeV mark. 

\subsection{Sensitivity}

\begin{figure}
    \centering
    
    \begin{overpic}[width=\textwidth, trim=0cm 0cm 0cm 0cm, clip]{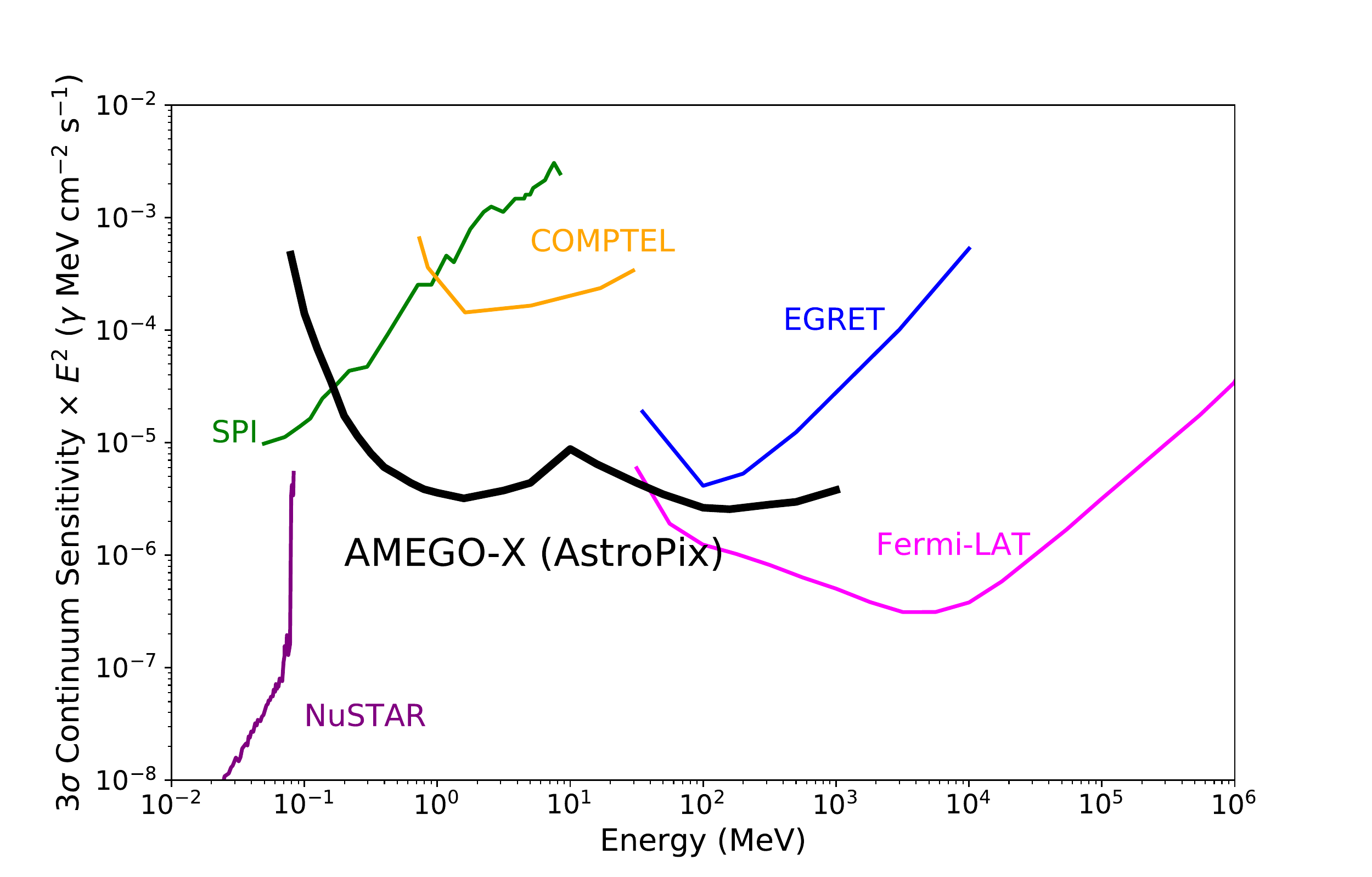}
        \put (15,53) {\textit{Preliminary}}
    \end{overpic}
    \caption{Predicted sensitivity for a 3-year all-sky survey for AMEGO-X with the AstroPix tracker, 0.5 mm pixel size, assuming an average off-axis angle of 37\degree{}, compared to other instruments.}
    \label{fig:sensi}
\end{figure}

We use the signal-to-noise ratio $SN = \frac{S}{\sqrt{S+B}}$ to evaluate whether or not a given source or source candidate is detected above the background fluctuations. Here, $S$ is the number of signal events and $B$ is the number of background events. The number of events is typically evaluated after some cuts have been applied on $\Delta \theta_C$ (for Compton events) or $\Delta \Theta$ (for pair events), where now the source position takes the place of the simulated gamma-ray direction, to select events that are compatible with having  been emitted by a given source. 

The sensitivity is defined as the lowest flux level detectable at the $3\sigma$ level (meaning, $SN\geq 3$) in a given energy bin. Figure \ref{fig:sensi} shows the expected sensitivity for a 3-year all-sky survey with AMEGO-X compared to other instruments. AMEGO-X will conduct the most sensitive survey as of yet in the energy range from hundreds of keV to tens of MeV, and will bridge the gap between hard X-rays and high-energy gamma rays.

\section{The AMEGO-X Science Program}

AMEGO-X aims to be a part of the global multi-messenger effort. Its main science goals are understanding the physical processes in both the extreme conditions around compact objects such as those involved in gravitational wave events and astrophysical jets seen in gamma-ray bursts, active galactic nuclei, and other extreme environments.

AMEGO-X will work as a survey instrument, orbiting the earth roughly once every 90 minutes and covering the entire sky every few orbits. AMEGO-X will search for the enhanced gamma-ray rates associated with long and short gamma-ray bursts. On short time scales (1 s), AMEGO-X will be sensitive to bursts with integrated fluxes between 120 keV and 1 MeV of $\geq$1.5 ph/cm$^2$/s (sensitivity requirement) or better. AMEGO-X is expected to detect hundreds of short GRBs every year, of which about 40-60 are expected to be detected with high significance ($\geq 8 \sigma$), good localization (68\% uncertainty radius of 1\degree{} or better) and a resolved energy spectrum. AMEGO-X will provide burst alerts (based on on-board event reconstruction and localization) to the community within 30 s or less for the majority of bursts to provide rapid follow-up observations.

AMEGO-X will monitor known active Galaxies for flares on timescales from hours to months, as well as searching for new MeV blazars. Gamma-ray flares coinciding with detections of an enhanced neutrino flux would be a smoking gun for hadronic acceleration in those galaxies. Spectral measurements in the MeV energy range are also crucial to distinguish between hadronic and leptonic emission models.

Other sources of interest for AMEGO-X are galactic particle accelerators such as pulsars, magnetars, and supernova remnants. 

Finally, as AMEGO-X will perform the most sensitive sky survey in MeV gamma-rays so far, we ``expect the unexpected'' and look forward to any surprised the Cosmos has in store for us!

\section{Status and Outlook}

We are aiming to propose AMEGO-X to the next MIDEX call for proposals, expected to come out in summer or fall 2021. In the meantime, multiple projects are going on to mature and test various detector components. For instance, the performance of the AstroPix pixel chips is being characterized at GSFC and KIT\citep{brewer2021astropix, astroPixNim}. ComPair\citep{amegoSPIE}, a prototype for the AMEGO detector, is slated for beam testing later this year and for a balloon test flight later on.


\bibliography{bib}

\section*{Acknowledgements}

The material is based upon work supported by NASA under award number 80GSFC21M0002. Many thanks to the rest of the AMEGO-X team!


%
%
%

\end{document}